\documentclass[preprint,aps]{revtex4}
\usepackage{graphicx}
\usepackage{booktabs}
\usepackage[usenames, dvipsnames]{color}
\usepackage{multirow}
\usepackage{tabularx}
\usepackage{setspace}
\usepackage{array,booktabs}
\usepackage{rotating}
\usepackage[section]{placeins}
\usepackage[ pdftex, plainpages = false, pdfpagelabels, 
                 pdfpagelayout = useoutlines,
                 bookmarks,
                 bookmarksopen = true,
                 bookmarksnumbered = true,
                 breaklinks = true,
                 linktocpage=all,
                 pagebackref=false,
                 colorlinks = true,
                 linkcolor = BrickRed,
                 urlcolor  = blue,
                 citecolor = BrickRed,
                 anchorcolor = green,
                 hyperindex = true,
                 hyperfigures
                 ]{hyperref} 
                 
\newcommand{\comment}[1]{}

\newcolumntype{L}[1]{>{\raggedright\arraybackslash}p{#1}}
\newcolumntype{C}[1]{>{\centering\arraybackslash}p{#1}}
\newcolumntype{R}[1]{>{\raggedleft\arraybackslash}p{#1}}

\begin{document}

\title{\href{http://www.necsi.edu/research/social/}{Visualizing the ``Heartbeat'' of a City with Tweets}} 
\author{Urbano Fran\c{c}a$^1$, Hiroki Sayama$^{1,2}$, Colin McSwiggen$^1$, Roozbeh
  Daneshvar$^1$ and \href{http://necsi.edu/faculty/bar-yam.html}{Yaneer Bar-Yam}$^1$}
\affiliation{$^1$ \href{http://www.necsi.edu}{New England Complex Systems Institute}\\  
238 Main St. Suite 319 Cambridge MA 02142, USA\\ 
$^2$ Binghamton University, State University of New York\\P.O. Box 6000, Binghamton, NY 13902-6000, USA}
\date{In final form May 21, 2014, Released Nov. 4, 2014}

\begin{abstract}
Describing the dynamics of a city is a crucial step to both understanding the human activity
in urban environments  and to planning and designing cities accordingly. 
Here we describe the collective dynamics of New York City and surrounding areas 
as seen through the lens of Twitter usage. In particular, we observe and quantify the patterns that 
emerge naturally from the hourly activities in different areas of New York City, and discuss
how they can be used to understand the urban areas.
Using a dataset that includes more than 6  million geolocated Twitter messages we construct
a movie of the geographic density of tweets. 
We observe the diurnal ``heartbeat'' of the NYC area. The largest scale 
dynamics are the waking and sleeping cycle and commuting from residential communities 
to office areas in Manhattan. Hourly dynamics reflect the interplay of commuting, 
work and leisure, including whether people are preoccupied with other activities or actively using Twitter. Differences between weekday and weekend dynamics point to changes 
in when people wake and sleep, and engage in social activities. 
We show that by measuring the average distances to the heart of the city one can 
quantify the weekly differences and the shift in behavior during weekends.
We also identify locations and times of high Twitter activity that occur because 
of specific activities. These include early morning high levels of traffic as people arrive and wait at air 
 transportation hubs, and on Sunday at the Meadowlands Sports Complex and Statue of Liberty. We analyze the 
role of particular individuals where they have large impacts on overall Twitter activity. Our analysis 
points to the opportunity to develop insight into both geographic social dynamics and 
attention through social media analysis.
\end{abstract}

\maketitle

\section{Introduction}

Identifying the universal principles and patterns that shape cities and urban human activities is crucial for both fundamental and 
practical reasons, as it not only sheds light on social organizations, but is also crucial for planning and 
designing safer and more sustainable cities. Complex systems approaches \cite{DCS} provide us with tools 
and new ways of thinking about urban areas and the ways they may correspond to living organisms \cite{Batty11}. The activities of people in urban areas are responsible for the emergence of patterns on large scales that define the dynamics of cities.

Until recently, one of the main obstacles to quantify such ideas 
was the lack of large scale data on flows of people and their activities. However, 
in the last decade there has been a surge of new technologies that make it possible to obtain 
real-time data about populations, and these new ``social microscopes'' have changed in fundamental ways the 
study of social systems \cite{Lazer09}.
Recent  examples include characterizing \cite{Gonzalez08, Pentland09} 
and predicting \cite{Song10, Sadilek12} individual human mobility patterns using mobile phone and 
social media data.  Toward that end, Twitter 
has been a valuable tool to 
track and to identify patterns of mobility and activity, especially using geolocated tweets. 
Geolocated tweets typically use the Global Positioning System (GPS) tracking capability installed on mobile devices 
when enabled by the user to give his or her precise location (latitude and longitude).  
Geolocated tweets have recently been used to study global mobility 
patterns \cite{Hawelka13} and spatial patterns and dynamics of sentiment \cite{Bertrand13, Frank13}.

Here we describe the collective dynamics of the greater metropolitan area of New York City (NYC) as reflected 
in the geographic dynamics of Twitter usage. We observe and quantify the patterns that 
emerge naturally from the hourly activities at different subareas, and discuss
how they can be used to understand the social dynamics of urban areas.
Twitter data can be understood not just by considering where people are but also by the extent
to which they are preoccupied or have time and attention to devote to Twitter posting. 
We collected more than 6 million geolocated messages from Twitter's streaming Application 
Programming Interface (API) \cite{API} from which more than 90$\%$  of geolocated tweets 
can be downloaded as they occur \cite{Morstatter13,Morstatter14}. From this data we observe 
wake/sleep cycles and the daily social ``heartbeat'' of the NYC area, reflecting the commuting 
dynamics from home to work in the diurnal cycle. We identify differences in weekday and 
weekend dynamics, and find specific locations where activity occurs at certain hours, 
including the early morning at air 
transit hubs. We also find anomalous events 
associated with specific individuals whose high engagement with Twitter at specific times 
can dominate their local region. We discuss how this dataset reflects the collective patterns 
of human activity and attention in both space and time.

\section{The dynamics of Twitter usage}

\begin{figure}
\centering
\includegraphics[width=1.0\textwidth]{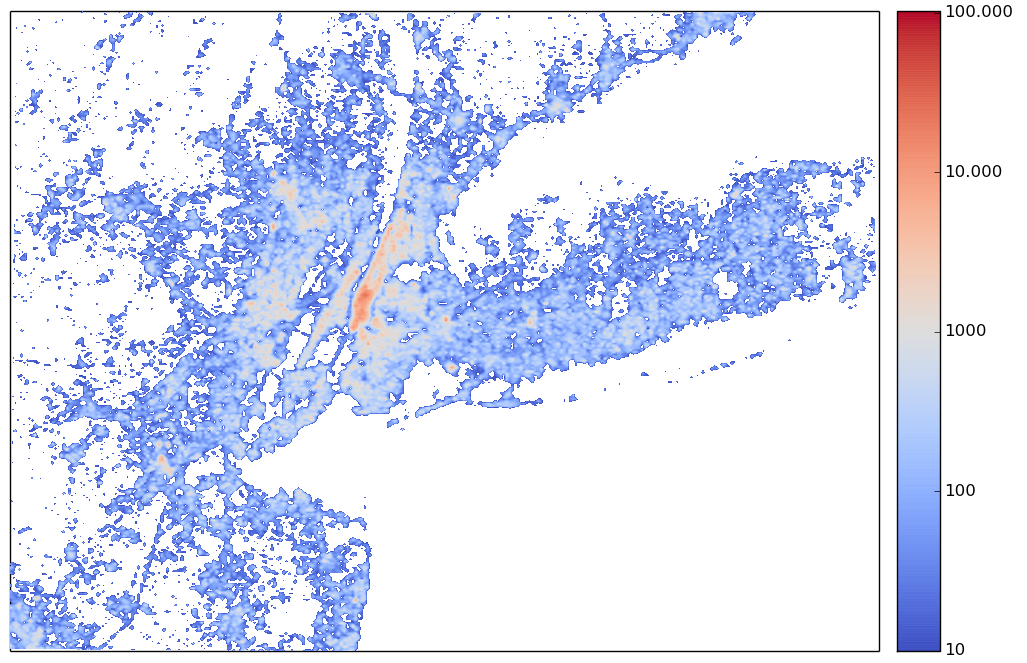}
 \caption{\label{fig:map} NYC geographical region with the locations of 6 million tweets shown. The sharp land-sea boundary is apparent as is the boundary of land area with high population density.}
\end{figure}

\begin{sidewaysfigure}
\begin{minipage}{1.0\textwidth}
\centering
\includegraphics[width=1.0\textwidth]{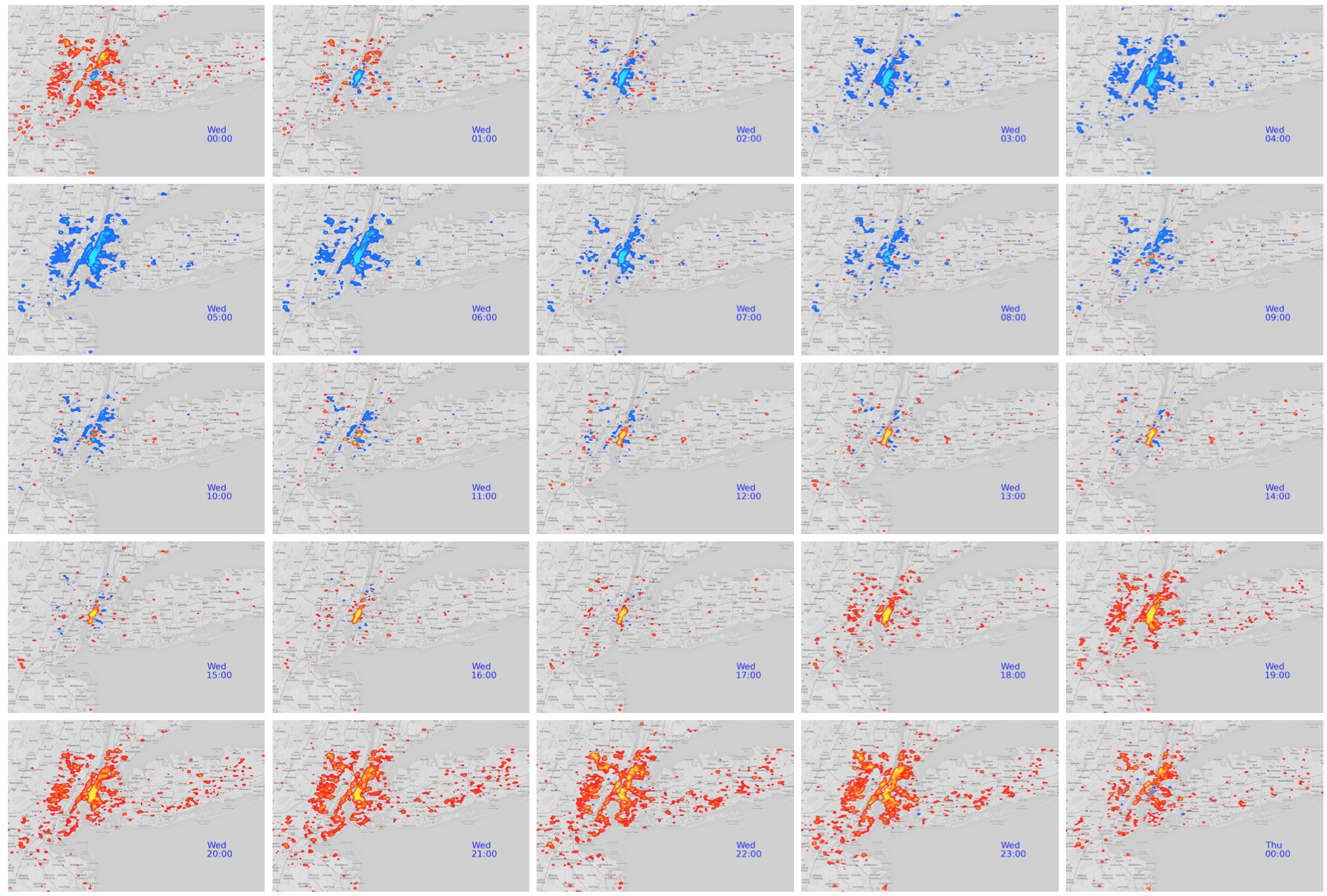}\\
\end{minipage}
\caption{\label{fig:day} Daily Twitter activity pattern of the NYC area for 24 hours of a ``typical'' weekday. The difference from the average number of tweets is shown revealing the times activity is higher (progressively red and yellow) or lower (progressively dark and light blue) than average for that location by more than a threshold amount $d_0=0.2$.}
\end{sidewaysfigure}

We collected tweets between the latitudes $[40.2^{\mathrm{o}}N, 41.2^{\mathrm{o}}N]$ and longitudes $[74.8^{\mathrm{o}}W, 
72.8^{\mathrm{o}}W]$ from August 19, 2013 to December 31, 2013. Fig. \ref{fig:map} shows the geographical Twitter coverage. We aggregated the data of the corresponding days of the week, and in hourly units of time, resulting in $7 \times 24 = 168$ time slices describing each hour of a ``typical week.'' We divided the geographic area into $90,000$ cells ($300 \times 300$). For each hour of the week we obtained the difference in that cell from the average number of tweets over the week as
\begin{equation}
d^i_{\mathrm{hour}} = \tanh\left[ \alpha ( n^i_{\mathrm{hour}} - \bar{n}^i) \right]
\end{equation}
where $n^i_{\mathrm{hour}}$ is the number of tweets in the cell $i$ at a given hour, 
$\bar{n}^i$ the average number of tweets in a given cell averaged over all days and hours, and $\alpha$ a
constant that controls the slope of the hyperbolic tangent ($\alpha = 0.04$ in all figures). Note that the 
$\tanh$ function bounds $d^i_{\mathrm{hour}}$ to the range $[-1, +1]$.
The values of $d^i_{\mathrm{hour}}$ were then used to generate a heat map of Twitter activity in the NYC area for a particular hour (see Fig. \ref{fig:day}). We also constructed two and three-dimensional movies of those patterns that can be accessed at \url{http://www.necsi.edu}.

The dynamics of Twitter can be understood first by recognition of the dominant diurnal wake-sleep cycle and geographic commuting to work in Manhattan from surrounding bedroom communities for the conventional workday hours approximately 9:00AM - 5:00PM. Earlier in the morning people tweet from their homes, and therefore Manhattan has much fewer tweets than average. Tweets are concentrated in Manhattan in the morning work hours and peak there at mid-day (around 13h), and become much more widely dispersed after work hours. The bedroom community activity is high in the evening, clearly visible at 10:00PM, and decreases as people go to sleep. 

\begin{figure}
\begin{minipage}{0.49\textwidth}
\centering
\includegraphics[width=1.0\textwidth]{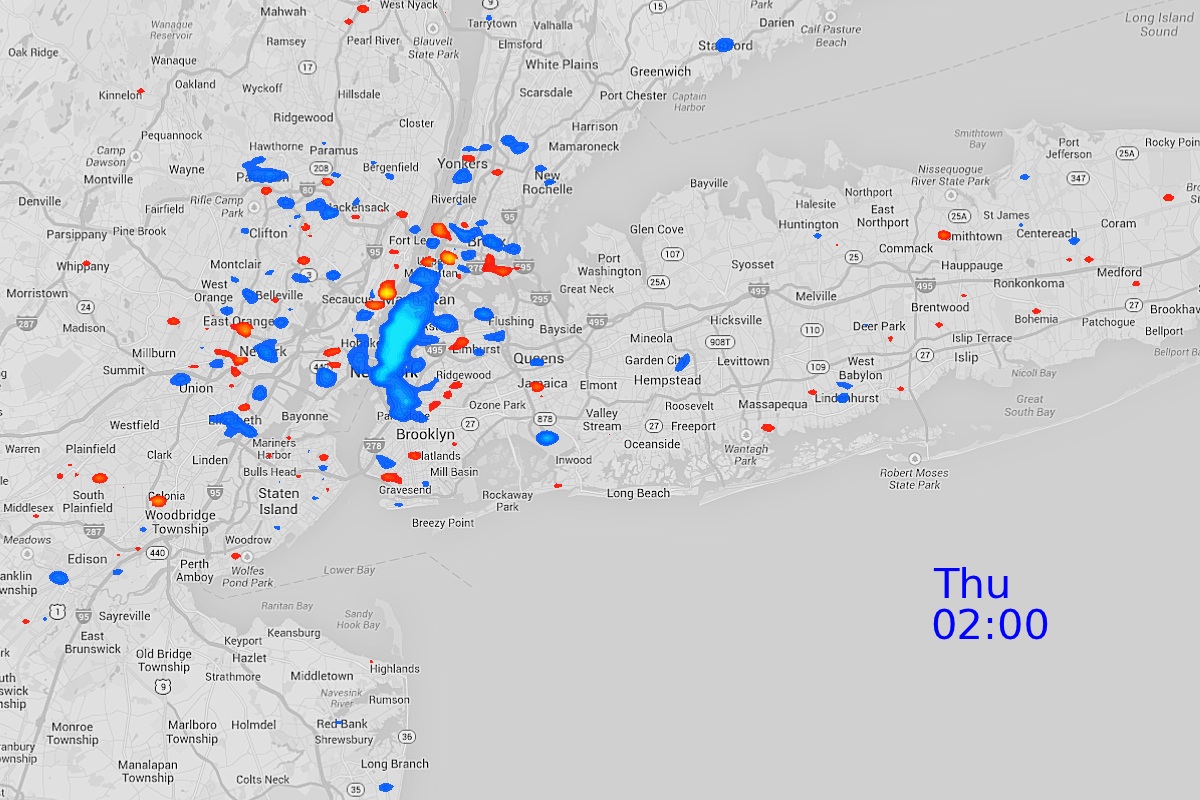}\\
\vspace{0.1cm}
\includegraphics[width=1.0\textwidth]{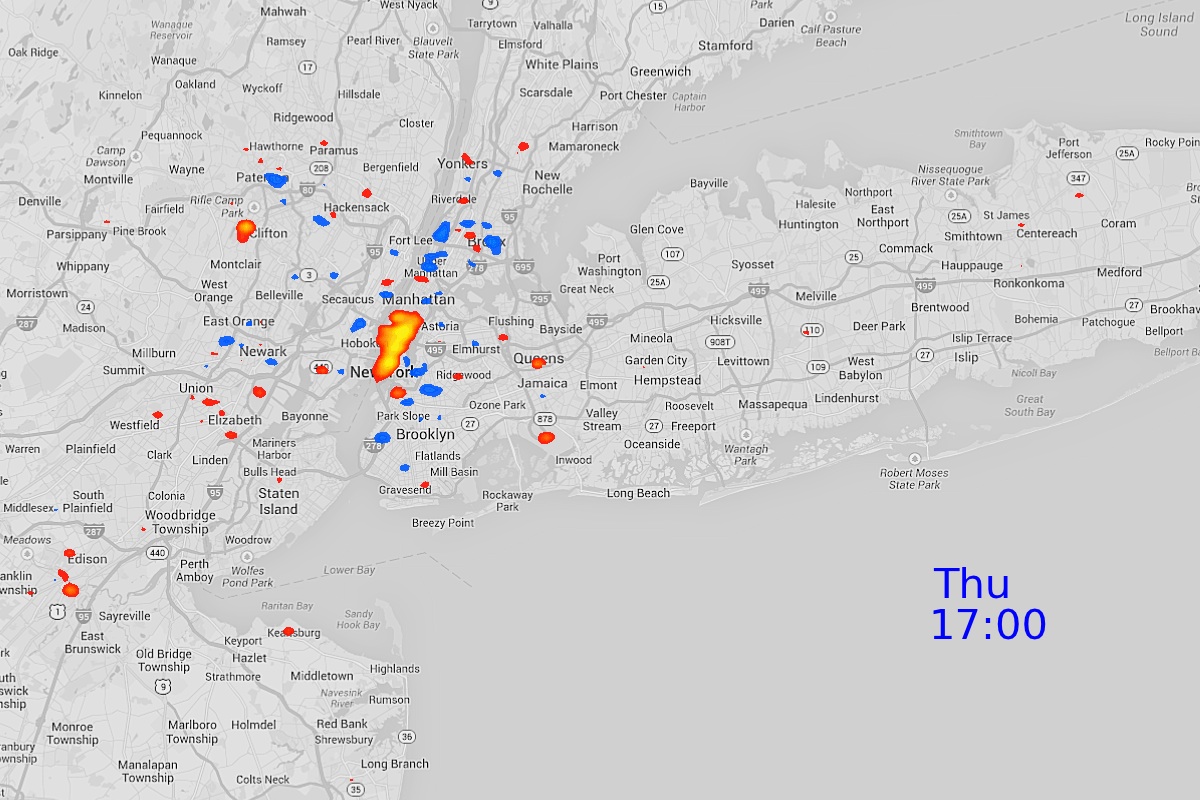}
\end{minipage}
\begin{minipage}{0.49\textwidth}
\includegraphics[width=1.0\textwidth]{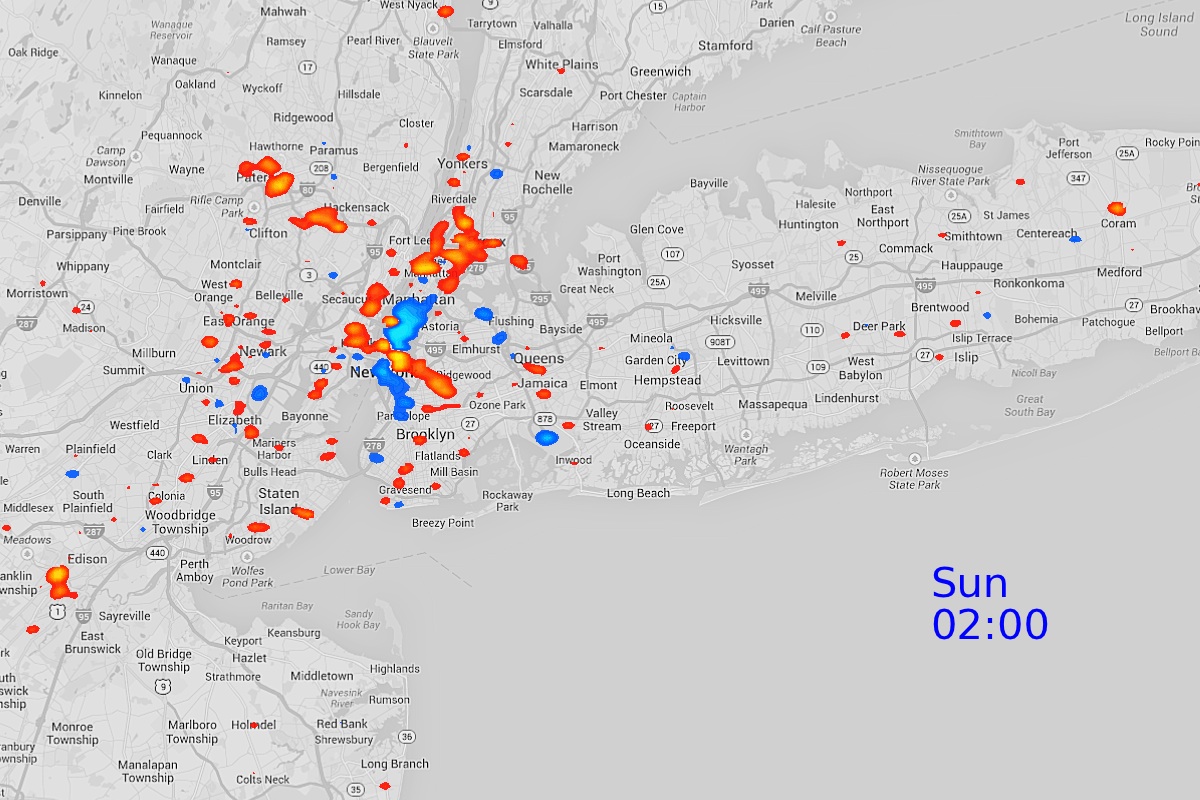}\\
\vspace{0.1cm}
\includegraphics[width=1.0\textwidth]{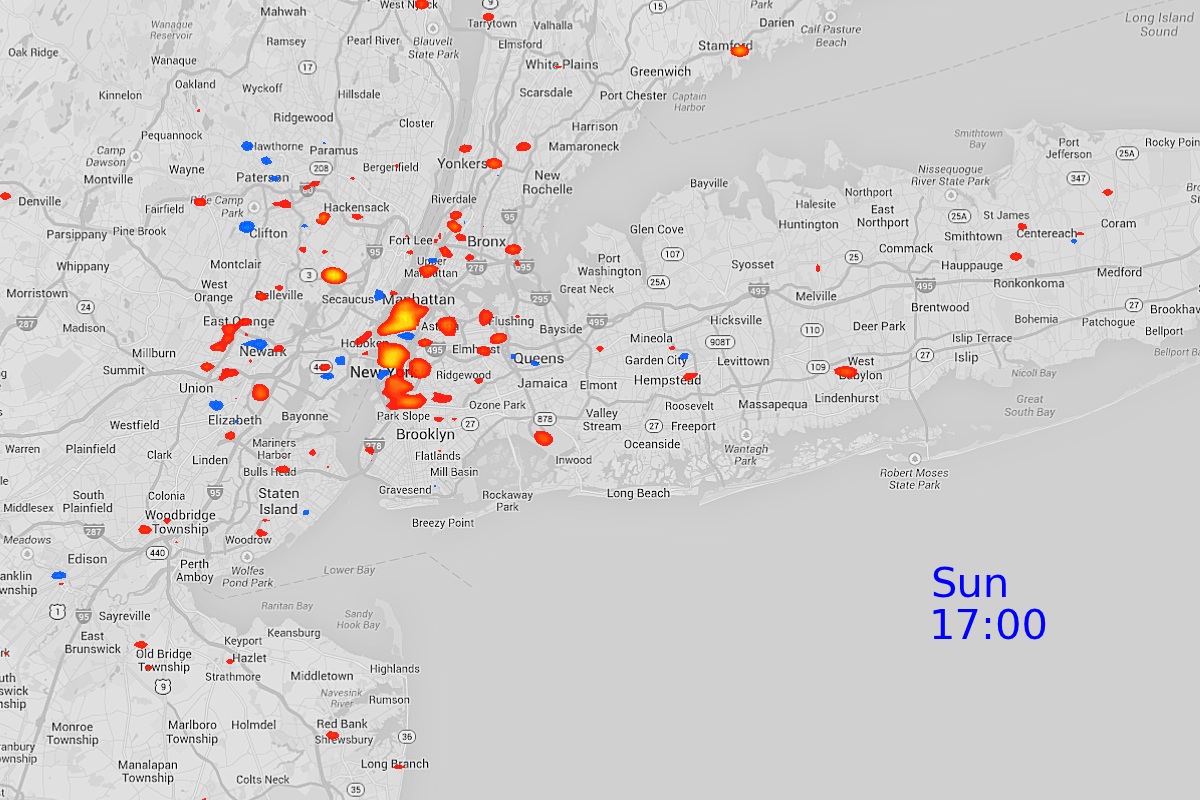}
\end{minipage}
\caption{\label{fig:weekdays} Twitter activity of weekdays compared to weekends, where the difference in the urban life can be clearly seen during the late night (top panels) and late afternoon (bottom panels). Colors as in Fig.  \ref{fig:day}}
\end{figure}

In addition to these largest scale daily patterns, patterns emerge when comparing 
weekdays (i.e., Monday to Friday) to weekends. In Fig. \ref{fig:weekdays} we compare 
the late-night and late-afternoon Twitter activity between Sunday and a workday at 2:00AM and 5:00PM.
While the workday activity is suppressed almost everywhere at 2:00AM, the nightlife activity at 2:00AM
Sunday morning (late Saturday night) has a unique pattern, with high activity in wide swaths of the city 
and suburbs. High levels of activity span a band extending from Lower 
Manhattan across to Brooklyn and Hoboken, New Jersey. Other high spots of night activity include the 
Bronx, and Union City and more specific spots in surrounding communities. 
Sunday afternoons also present an unusual pattern of widely dispersed but localized spots of activity likely corresponding to tweets in residential community areas of activity. 
Moreover, a peak of activity is observed in Central Park most of the day Sunday that is not 
observed on other days of the week. 

\begin{figure}
\centering
\begin{minipage}{0.48\textwidth}
\includegraphics[width=1.0\textwidth]{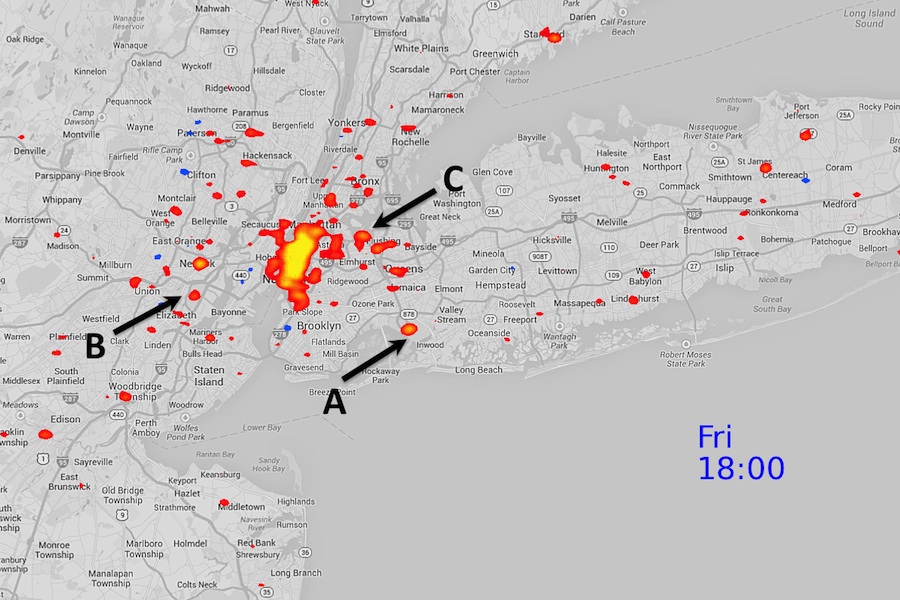}\\
\vspace{0.1cm}
\includegraphics[width=1.0\textwidth]{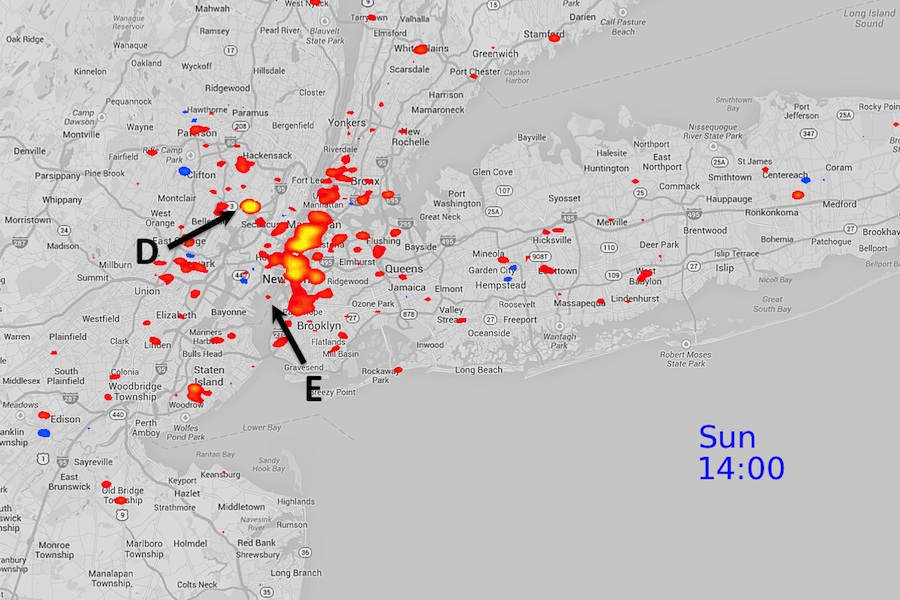}
\end{minipage}
\begin{minipage}{0.46\textwidth}
\vspace{0.5cm}
\centering
\includegraphics[width=1.0\textwidth]{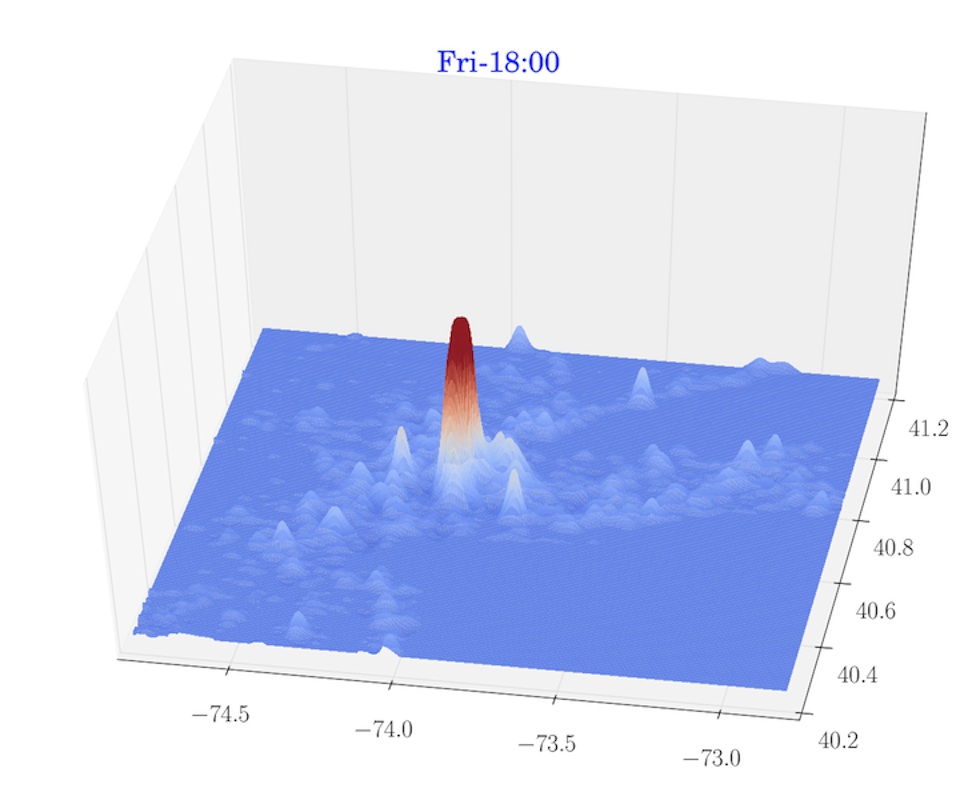}\\
\includegraphics[width=1.0\textwidth]{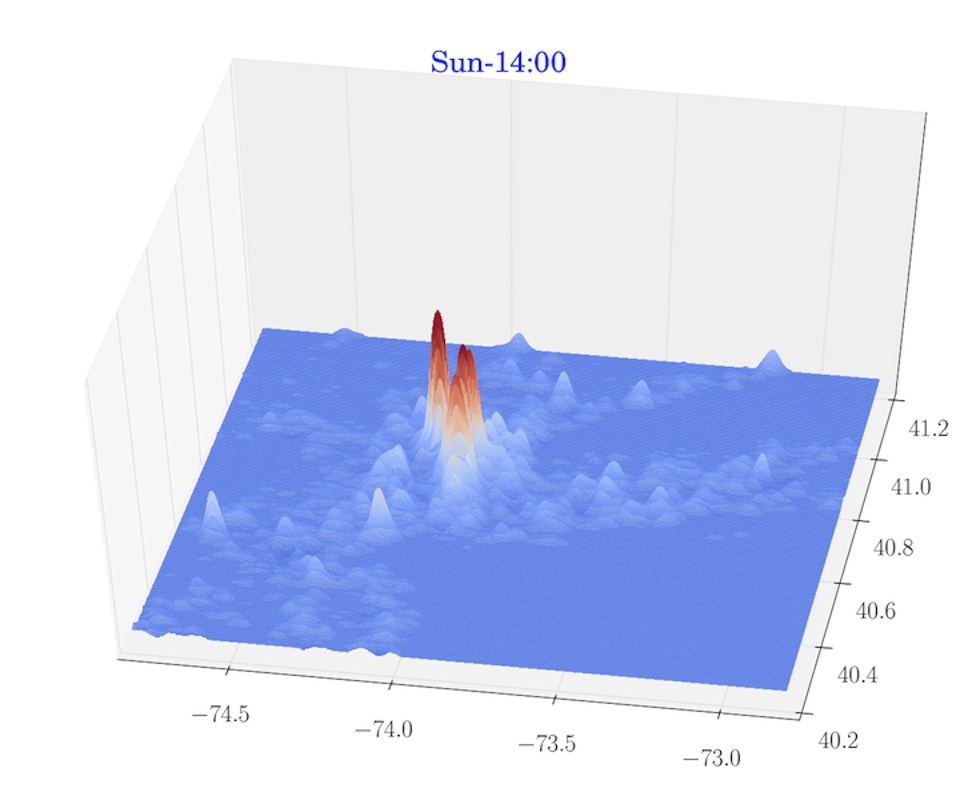}
\end{minipage}
\caption{\label{fig:spots} Examples of times of high activity at locations listed in Table \ref{tab:spots}. The left panel shows annotated activity plots using the colors as in Fig. \ref{fig:day}, and the right panel shows the corresponding height as a three dimensional surface.}
\end{figure}

We also find other interesting activities at particular locations and times, as can be seen in Fig. 
\ref{fig:spots}. The top row shows the high activity at 6:00PM on Friday at the three main Airports of the  
area: John F.~Kennedy (JFK) airport ({\bf A})  \cite{JFK}, Newark Airport ({\bf B}) \cite{Newark} 
and La Guardia Airport ({\bf C}) \cite{LG}.  
Weekend activities are seen in the bottom row, we point out 
the Meadowlands Sports Complex  ({\bf D}) \cite{Mead} and the Statue of 
Liberty  ({\bf E}) \cite{Liberty}.

\begin{table}[ht]
\centering
\caption{Examples of high-activity areas}
\begin{tabular}{|L{3.0in} L{3.5in}|}
\hline 
{\bf Location} & {\bf Typical Peak hours}  \\
\hline 
{\bf A}. JFK Airport & Mon-Sat 6AM; Everyday 4PM-6PM; Sun 8PM\\
{\bf B}. Newark Airport & Everyday 4PM-7PM \\
{\bf C}. La Guardia Airport & Sun-Fri 4PM-7PM\\
{\bf D}. Meadowlands Sports Complex & Sat 2PM-9PM; Sun 9AM-7PM \\
{\bf E}. Statue of Liberty & Sat 11AM-4PM; Sun 2PM \\
\hline 
\end{tabular}
\label{tab:spots}
\end{table}

\begin{figure}
\centering
\begin{minipage}{0.48\textwidth}
\includegraphics[width=1.0\textwidth]{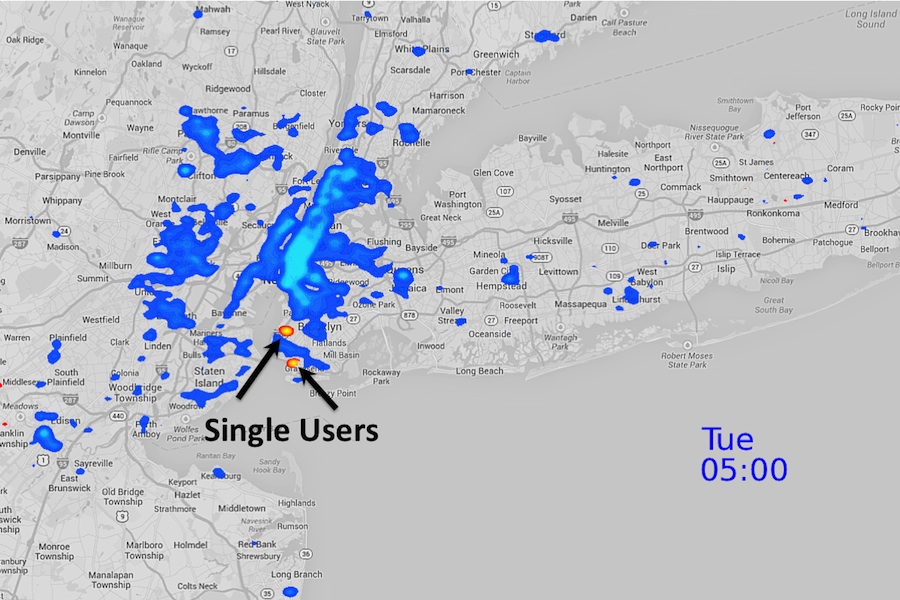}\\
\end{minipage}
\begin{minipage}{0.48\textwidth}
\centering
\includegraphics[width=1.0\textwidth]{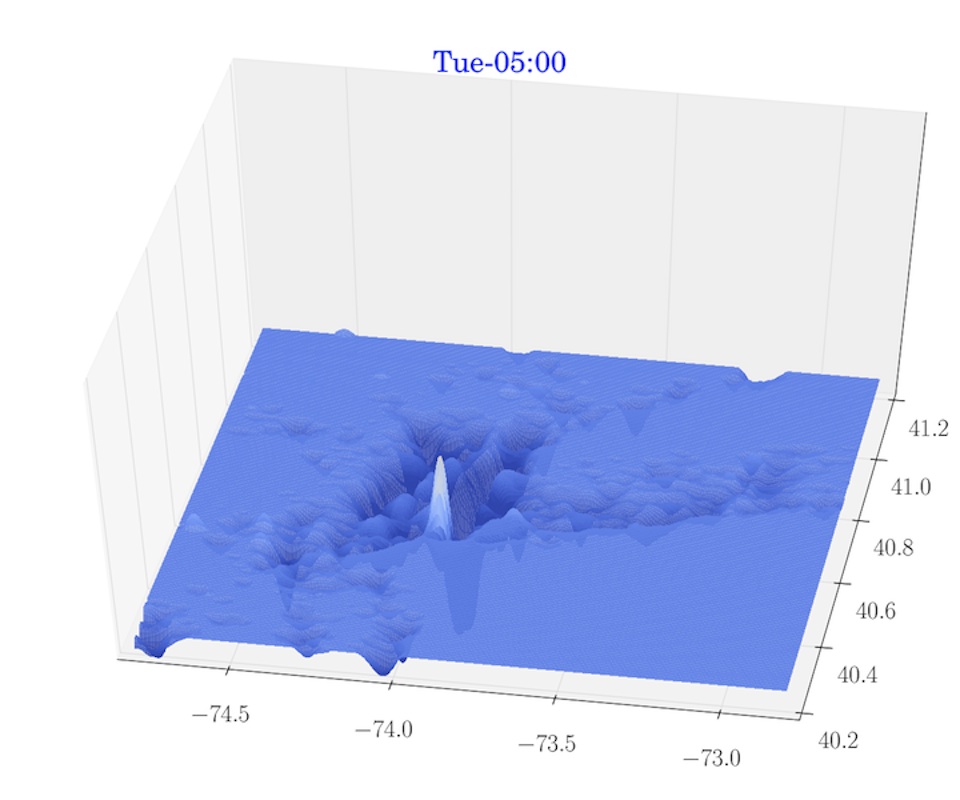}\\
\end{minipage}
\caption{\label{fig:anom} Single user anomaly: the annotated peak corresponds to a single 
user tweeting more than 10 times the average of other users.}
\end{figure}

Finally, Fig. \ref{fig:anom} shows examples where an individual's behavior dominates the collective twitter activity. We used statistical measures to discover that a single Twitter user was responsible for most tweets in a specific region around 5:00AM on weekdays. Detecting such anomalies enables distinguishing collective social dynamics from individual behavior that can at times dominate aggregate measures.

\begin{figure}[b]
\centering
\includegraphics[width=1.0\textwidth]{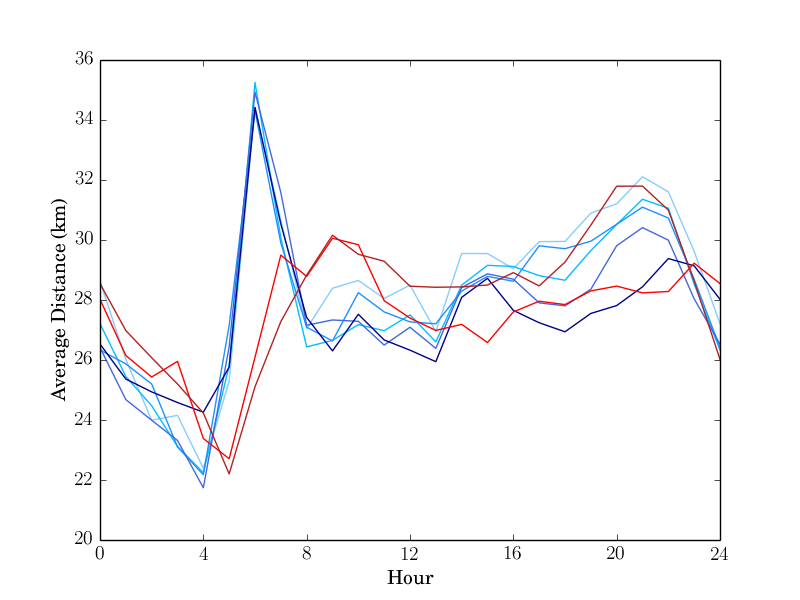}
 \caption{ \label{fig:avgd}
Average distance from Central Park at the different hours of the day for each day of the week. The working days of the week, Monday to Friday, are shown in different shades of blue (light blue for Monday and the darkest blue for Friday), while the weekends are shown in shades of red (red for Saturday and dark red for Sunday).}
\end{figure}

To identify a single measure that can capture key aspects of the collective behavior of the city we considered the average distance of the tweets from a location in Manhattan (Central Park, $40^{\mathrm{o}} 46' 43.35''N$, $-73^{\mathrm{o}} 58' 5.27''W$) over time. The distance was calculated using the Haversine formula \cite{Meyer10},
\begin{equation} \label{eq:Hav}
d = 2 r_{\oplus} \arcsin \left[ \sqrt{\sin^2\left( \frac{\phi_2 - \phi_1}{2} \right) +
\cos(\phi_1)\cos(\phi_2) \sin^2 \left( \frac{\lambda_2 - \lambda_1}{2} \right) } \right] \ , 
\end{equation}  
where $r_{\oplus} = 6367$ km is the average radius of the Earth, $\phi_i$ and $\lambda_i$, ($i =1,2$) 
are the latitude and longitude of the point $i$. The results are shown in Fig. \ref{fig:avgd}. 

We find that while many people are expected to be located in bedroom communities, the average 
distance to Manhattan falls overnight, reflecting the fact that downtown activity continues through the
night so that most Twitter activity at late hours/early morning (12:00AM to 5:00AM) happens in areas in or close 
to Manhattan. During weekdays a sharp peak at 6:00AM corresponds to people tweeting in suburban areas 
before commuting. After 8:00AM the distance grows gradually during the day and through the evening, until 
it falls again overnight. A change in this pattern is observed during weekends. There is a horizontal shift in 
the curves for Saturday and Sunday of about 3 hours, reflecting people's tendency to tweet later into 
the night and wake up later the next morning. The pre-commuting peak is entirely missing.

\section{Conclusions}

In this paper we characterized, for the first time, the patterns of weekly 
activity of the NYC area using more than 6 
million geolocated tweets posted between Aug 19, 2013 and Dec 31, 2013.  
We related the collective geographical and temporal patterns of Twitter usage to
activities of the urban life and the daily ``heartbeat'' of a city.  The largest scale 
daily dynamics are the waking and sleeping cycle and commuting from the suburbs
to office areas in Manhattan, while the hourly dynamics reflects the interplay between commuting, 
work and leisure. 

We showed not only that Twitter can provide 
insight into understanding human social activity patterns, but also that our analysis is 
capable of identifying interesting locations of activity by focusing on departures from global behaviors. 
We observed a peak of Twitter usage in the suburbs of NYC before people begin their workday, and during evening hours, while the main activity during the workday and late at night is in Manhattan.

In addition to the daily differences, we also characterized the weekly patterns, especially differences between 
the weekend and weekdays. Daytime recreational activities concentrate at identifiable locations spread widely 
across residential areas of the city. 
We determined more specific times and locations of high activity at air transportation
hubs, tourist attractions and sports arenas.
We analyze the role of particular individuals where they have large impacts 
on overall Twitter activity, and find effects which may be considered outliers in discussion of social activity, 
but are an important aspect of human activity in the city. 

We explored the use of the average distance  from downtown to understand the dynamics of Twitter usage, and 
while weekdays are similar in the geography of Twitter usage, the sleep
waking cycle is shifted later by about 3 hours during the weekends, and a pre-commuting Twitter activity peak
is absent.  Taken together, those results demonstrate the potential of using social media analysis 
to develop insight into both geographic social dynamics and activities, and opens the possibility to 
understand and compare the life of cities on various scales.

\end{document}